\documentclass[5p]{elsarticle} %two columns 

\usepackage{lineno,hyperref}
\usepackage{footnote}
\modulolinenumbers[5]

\usepackage{graphicx}
\usepackage{mathcomp}
\usepackage{amsmath}
\usepackage{float}
\usepackage{xcolor}

\journal{Physica A}
\begin{document}

\begin{frontmatter}

\title{A statistical mechanical model for drug release: \\ relations between release parameters and porosity}

\author[gomesfaoaddress]{M\'arcio Sampaio Gomes-Filho}

\author[marcoaddress]{Marco Aur\'elio Alves Barbosa}
\ead{aureliobarbosa@unb.br}

\author[gomesfaoaddress]{Fernando Albuquerque Oliveira}
\ead{fao@fis.unb.br}

\address[gomesfaoaddress]{Instituto de F\'isica, Universidade de Bras\'ilia, Bras\'ilia-DF, Brazil}

\address[marcoaddress]{Programa de P\'os-Gradua\c{c}\~ao em Ci\^encia de Materiais, Faculdade UnB Planaltina, Universidade de Bras\'ilia, Planaltina-DF, Brazil}

  \begin{abstract}
    A lattice gas model is proposed for investigating the release of drug molecules on devices with semi-permeable, porous membranes in two and three dimensions. The kinetic of this model was obtained through the analytical solution of the three-dimension diffusion equation for systems without membrane and with Monte Carlo simulations. Pharmaceutical data from drug release is usually adjusted to the Weibull function, $\exp [-(t/\tau)^b ]$, also known as stretched exponential, and the dependence of adjusted parameters $b$ and $\tau$ is usually associated, in the pharmaceutical  literature, with physical mechanisms dominating the drug dynamics inside the capsule.  The relation of parameters $\tau$ and $b$ with porosity $\lambda$ are found to satisfy, a simple linear relation for between $\tau$ and $\lambda^{-1}$, which can be explained through simple physically based arguments, and a scaling relation between $b$ and $\lambda$, with the scaling coefficient proportional to the system dimension.
    %Our results can be used for estimating release patterns in real drug carriers, such as porous silicon nanoparticles, were porosity can be controlled by fabrication and/or surface functionalization.
  \end{abstract}


\end{frontmatter}

\section{Introduction}
The advances in the synthesis technology of porous materials allowed the development of new matrices (monolithic) and membrane pharmaceutical devices where size, shape and pore distribution can be fine controlled during the fabrication process~\cite{gultepe2010, jeon2012,  yazdi2015, siepmann2011:book}.  Understanding the connection between drug release rates and the characteristics of the device has an enormous potential for improving the treatment of various diseases. 

Mathematical modeling of drug release usually involves finding the proper form of a diffusion equation considering the essential physical phenomena occurring as particles diffuse through the capsule device~\cite{costa2001, siepmann2008, caccavo2019}. 
%and some comments about empirical models~\cite{costa2001, siepmann2008, siepmann2011b, caccavo2019}
In the pharmaceutical literature it is also a common procedure to fit drug release data to semi-empirical functions and use this information to obtain insights onto the processes through which the drug is released. Since many factors can contribute to determine the final drug release, this procedure can be subject to ambiguous interpretations that could make the data analysis even more confusing~\cite{siepmann2011b}. Thus, a more systematic approach for understanding the relation between release data and physical processes occurring inside the capsule is desired. We work in this direction by simulating minimalist lattice models devised to describe both drug and physical device (or capsule) and investigate the relation between drug release patterns and the system porosity through the semi-empirical parameters of the Weibull function.

In this work a lattice gas model is proposed for investigating the release of drug molecules encapsulated on devices with semi-permeable, porous membranes in two and three dimensions, following a previous work on 1D and 2D systems~\cite{gomesfilho2016}. Release patterns were obtained through analytical solution of the three-dimension diffusion equation, for systems without membrane, and Monte Carlo simulations (MC), for systems with porous membrane, and adjusted to the Weilbull function, $\exp [-(t/\tau)^b ]$, which is also known as stretched exponential. The dependence of the characteristic time $\tau$ with the membrane content, defined as the inverse power of porosity, $\zeta = \lambda^{-1}$, was found to satisfy linear relation, that is justified using reasonable physical arguments. The parameter $b$ was found to satisfy a scaling relation with $\zeta$ from a regime without membrane for up to 90\% of membrane coverage.

This article is organized as follows, in the next section we present the Weibull distribution and discuss previous investigations about its semi-empirical parameters using statistical mechanical models, the current model and the simulations protocol are introduced in Section~\ref{sec:model}, while our results and discussions are presented in Section~\ref{sec:results}. An analytical solution for the  diffusion equation of a continuous system similar to our 3D lattice model is presented in~\ref{appendixA}.

\begin{figure*}
  \begin{center}
    \includegraphics[scale=.9]{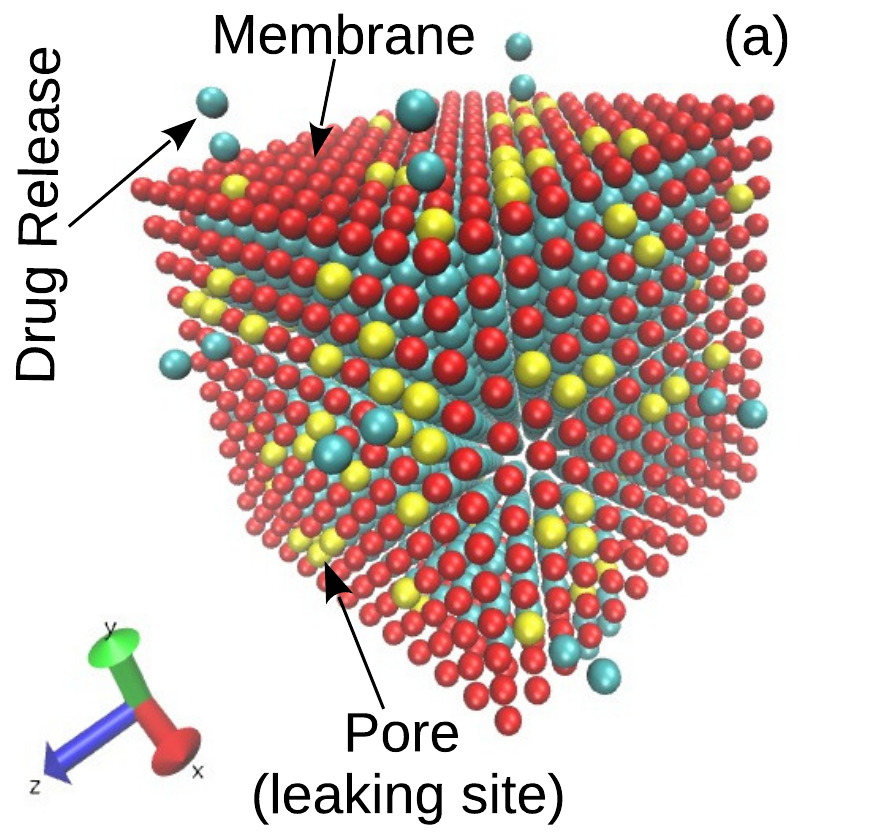}  \includegraphics[scale=.5]{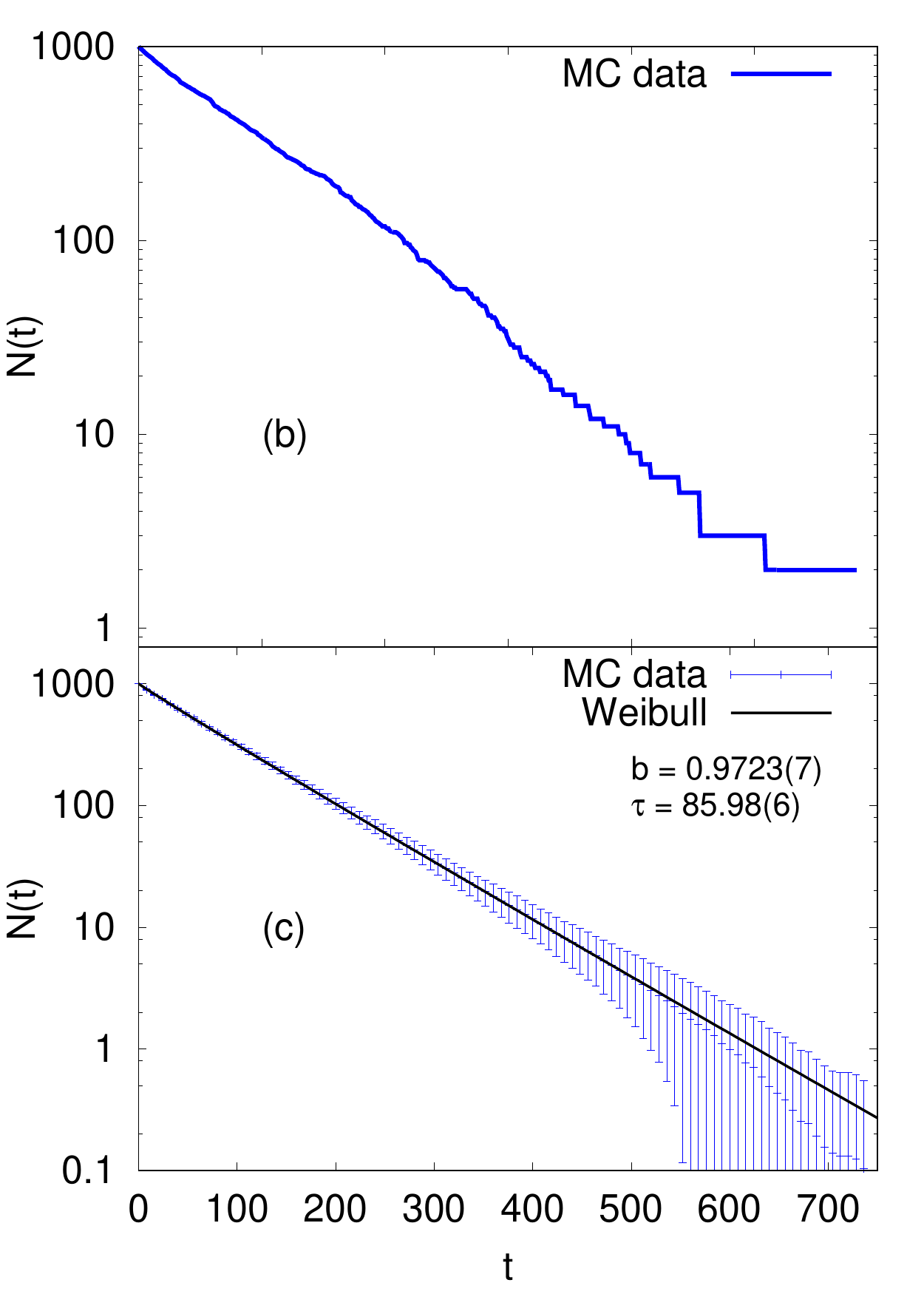}
    \caption{\label{fig:model} (Color online) Representation of a three-dimensional device  with size $L=10$ and porosity $\lambda = 1/6$ encapsulating drug molecules. A few molecules (blue) are escaping through pores (yellow), while most particles are kept inside the capsule by the membrane covering it (red). Release dynamics of a single simulation is illustrated in (b) while in (c) the same quantity is averaged over 250 different simulations.}
  \end{center}  
\end{figure*}

\subsection{Weibull distribution}
The Weibull distribution function was originally proposed by Waloddi Weibull in 1951 as the `simplest possible' empirical function that could be used to adjust non-linear experimental data from complex systems~\cite{weibull1951}. Distributions on systems as diverse as electric bulb duration, life expectancy in human populations, and yield strenght of steel where well interpolated in the original work. For further details on the Weibull distribution function see the reference~\cite{weibullbook}. On the pharmaceutical research it was first used to adjust data from drug release on 1972 by Langenbucher~\cite{langenbucher1972} and, since then, it became common to use the Weibull function to obtain phenomenological insights into the intrinsic mechanisms governing the drug release. As discussed by Slater~\cite{slater05:ijpharm}, it is appropriate to write this as:
\begin{equation}
  \frac{N(t)}{N_{0}} = \exp \left [- \left ( \frac{t}{\tau} \right )^{b} \right],
  \label{eq:weibull-1}
\end{equation}
where  $N(t)$ is the amount of drug molecules inside the device as a function of time $t$ and $N_0 \equiv N(0)$  is the initial number of drug  molecules inside the device. The parameter $\tau$ is associated with the time where approximately $63\%$ of the drug has been released and $b$ with the physical mechanisms leading to drug release within the device~\cite{langenbucher1972, kosmidis2003b, papadopoulou2006}.  
Expression~\ref{eq:weibull-1} is also known as stretched exponential, and $b\neq 1$ in diffusive models is usually associated to process with coloured noises and/or memory effects~\cite{vainstein2006,oliveira2019}.

Lattice models have been used to investigate the relation between the release and capsule parameters~\cite{burnette1984, singh2019}. While early works were devoted to finding scaling properties of regular, fractal and multi-channel matrices~\cite{bunde1985, balazs1985}, more recent work using lattice models (usually on the pharmaceutical literature) was focused on validating the usage of semi-empirical functions, as well as understanding the relation between various physical aspects of the capsule, such as shape, system dimension, drug concentration on drug release~\cite{kosmidis2003a, kosmidis2003b, dokoumetzidis2005, villalobos2006a, villalobos2006b, villalobos2009c,  papadopoulou2006, kosmidis2007, kosmidis2008, martinez2009, dokoumetzidis2011, hadjitheodorou2013, hadjitheodorou2014, kalosakas2015, gomesfilho2016, christidi2016,  kosmidis2018}. 
%In some of these works~\cite{bla}, the release profile was adjusted to Weibull function and mathematical relations between the semi-empirical parameters with key physical aspects were proposed~\cite{bla}. 
In the current work we investigate the relation between porosity and drug release by using the Weibull function as an interpolating function, as in the previously mentioned work, assuming that the semi-empirical parameters $b$ and $\tau$ retains relevant physical information about the nature of the capsule. We present physical arguments which result in a linear relation between characteristic release times and membrane content (which will be defined in terms of porosity, in the following) and, besides that, results from Monte Carlo Simulations indicate that the parameter $b$ is related to porosity through simple scaling relations.

\begin{figure*}
\begin{center}
  \includegraphics[scale=.68]{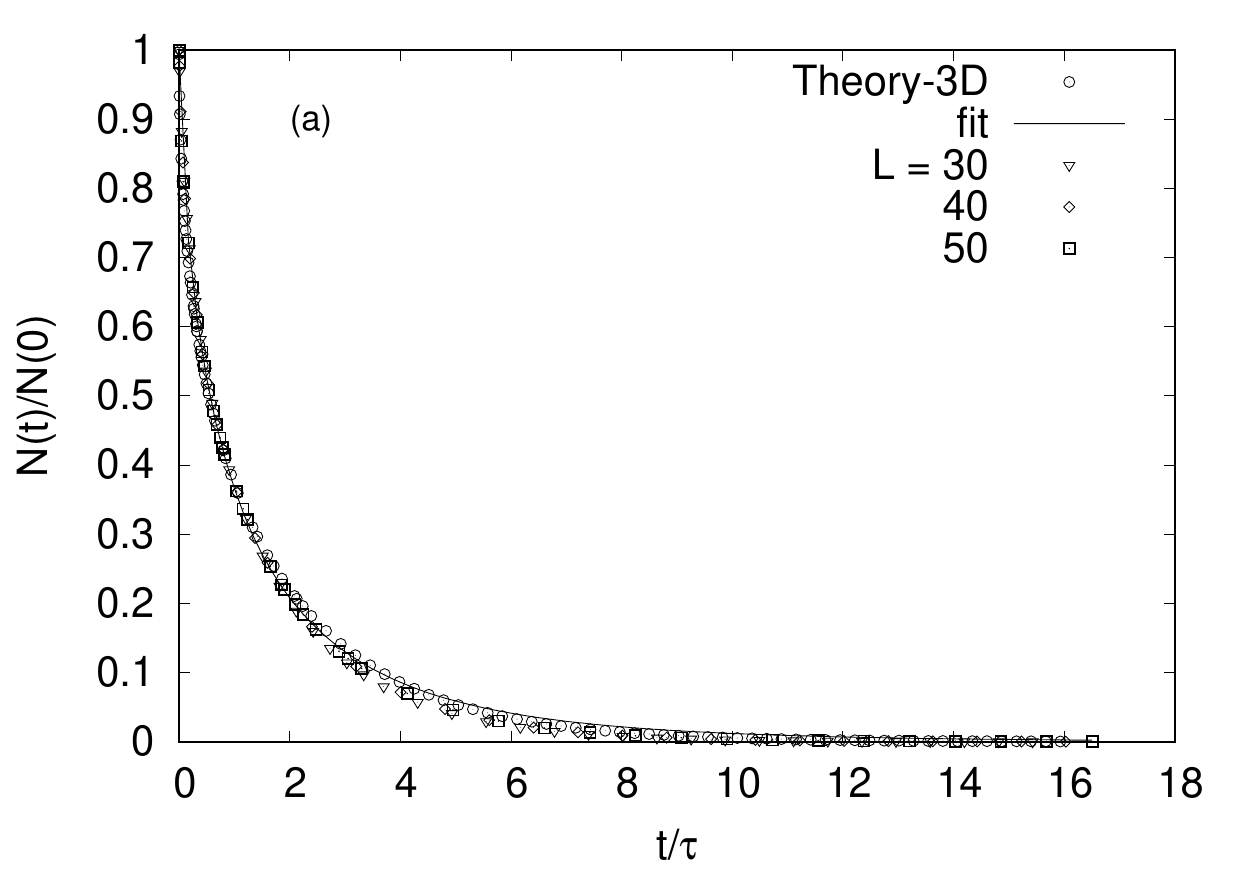}
  \includegraphics[scale=.68]{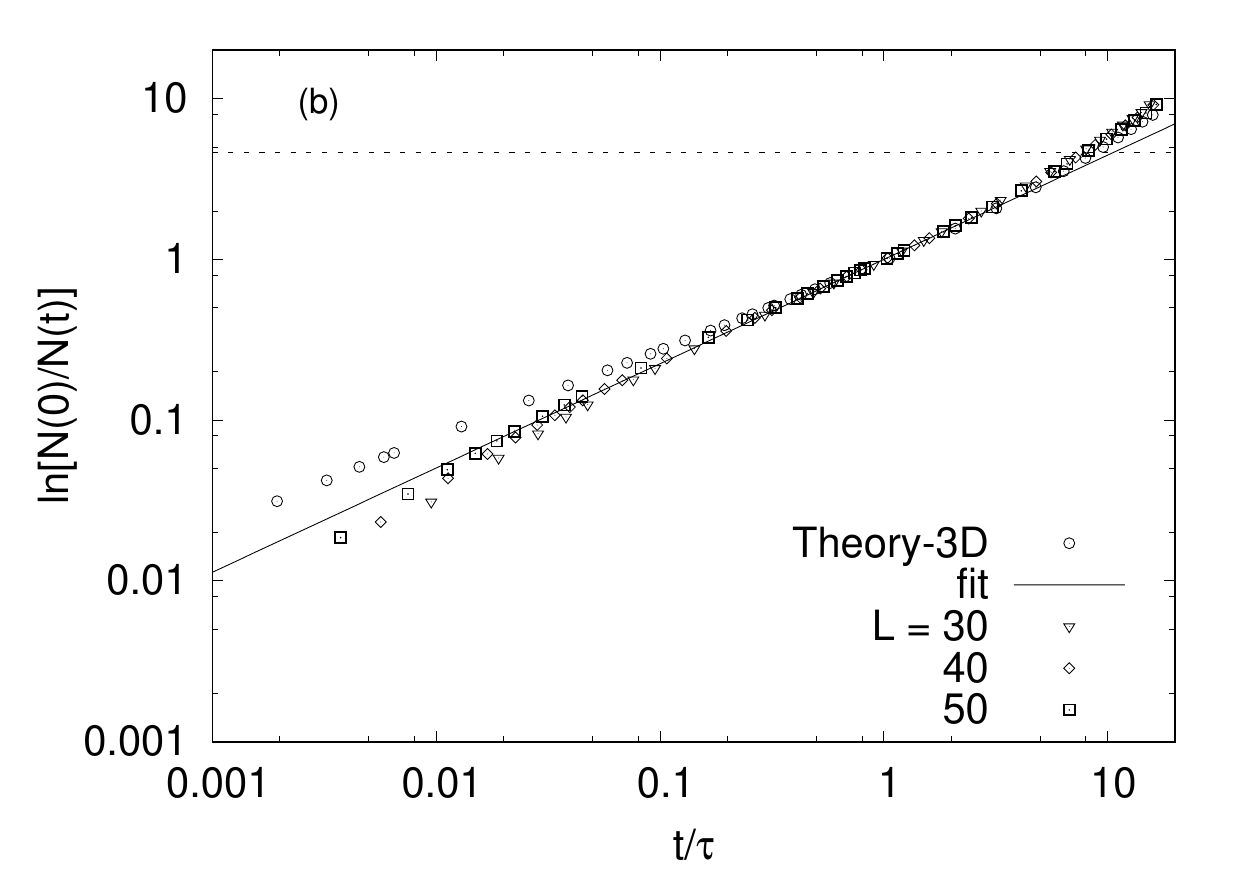}
  \caption{\label{fig:theory-fit} (a) For a cubic device without membrane, the Weibull function is fitted to release data from both the analytical solution of diffusion equation (points) and Monte Carlo simulations with different sizes.  In (b) the same data is shown in a triple log graph.}
\end{center}
\end{figure*}

\section{\label{sec:model}Model and Monte Carlo Simulation}

The model investigated here is based on a previous work on 1D and 2D system inspired on the lattice gas model to represent a system of device capsule plus drug, and simulate the non-equilibrium drug release process~\cite{gomesfilho2016}. The current modelling includes 2D square and 3D simple cubic lattice, both with size $L$, for representing the devices delivering drugs. Drug molecules are represented as single particles occupying lattice sites and kinetics is obtained by allowing particles to move randomly to unoccupied nearest neighbor sites. Excluded volume interaction precludes two drug molecules to share the same site. A membrane with edge size $L+1$ covers the device and acts by blocking drug molecules from leaking to the outside environment. There are $n$ randomly positioned leaking sites on this membrane ``surface'', and the porosity parameter $\lambda$ can be define as $\lambda_d=n/n_{S,d}$, where $n_{S,d}$, the number of surface sites on a capsule with dimension d, is identical to $4L$ ($n/6L^2$) for 2D (3D) system. A single 3D model device with $L=10$ ($L^3 = 10^3$ sites) and porosity $\lambda = 1/6$ ($100$ leaking sites) is represented on Fig.~\ref{fig:model} (a). It is important to mention that the current implementation of our simulation protocol fixes the code of our previous 2D simulations~\cite{gomesfilho2016}~\footnote{The code used for generating the MC simulations in our previous work~\cite{gomesfilho2016} was written in C and presented a bias in the procedure for sampling membrane pores, towards one of the edges. This error introduced an artifact in the behavior of $b$ as a function of $\lambda$, thus hiding the scaling relations that we observed here for both 2D and 3D systems.}.

Drug release kinetics is obtained through Monte Carlo simulations, as discussed in the literature~\cite{kosmidis2019, macheras2006, barat2006}, with the difference that both the average and standard deviation of particle numbers are collected on each time step. On this work, standard deviations will be used to weight data while adjusting the entire release curve to the Weibull function. The reader is referred to the original work for more details, but the main points of the MC simulation protocol are described here for clarity. The initial configuration of all simulations is a filled device, \textit{i.e.}, the initial number of particles is $L^2$ ($L^3$) on 2D (3D) and, on each device, its membrane leaking sites are randomly assigned in accordance with its porosity value $\lambda$. On each simulation drug particles are randomly selected and allowed to try a random jump to a nearest neighbor site, and it is allowed only
if the new site is empty. After each attempted movement time is incremented by a factor $\delta t/N$, where $\delta t$ is a standard MC time unit and $N$ is the number of particles remaining inside the capsule.  Whenever a particle jumps into a leaking site (pore) it is removed from the system, thus decreasing the number of particles inside the capsule. A number $\mathcal{R}_{L,\lambda}$ of different simulations with the same size and porosity run in parallel, each simulation ending when at least $99.99\%$ of the drug particles have leaked from the capsule. After finishing all simulations  release profiles are obtained by calculating the average number of particles, $N(t)$, and standard deviation $\sigma (t) = \sqrt{\langle N(t)^2 \rangle - \langle N(t) \rangle^2}$ through the $\mathcal{R}_{L,\lambda}$ simulations, with $\mathcal{R}_{L,\lambda}$ numbers being  equal to $1000$ and $250$  in 2D and 3D systems, respectively. 

A capsule with size $L=10$ and a membrane with $n=100$ leaking sites, corresponding to a porosity $\lambda=1/6$, was simulated to illustrate a typical model system, as shown in Figs.~\ref{fig:model}(a)-(c). A single configuration is illustrated in (a) while the output from a single simulation and the final drug release profile, as averaged over 250 simulations, are presented in Figs.~\ref{fig:model} (b) and (c).

\section{\label{sec:results} Results and discussion}

We present results from MC simulations of devices in two and three dimensions with edge sizes given by $L_{2D} \in \{150, 200\}$ and $L_{3D} \in \{30, 40, 50 \}$, respectively. The effect of porosity was  investigated by varying the parameter $\lambda$ from values corresponding to $1\%$ to $100\%$ of pores, with increment steps of $1\%$~\footnote{In both 2D and 3D models $\lambda$ values were in the range between 0.01 to 1.00 with increments of 0.01.}. Once a release curve is obtained, as illustrated in Fig.\ref{fig:model} (c), the Weibull function is adjusted to the data set and a scaling analysis is performed on the dependence of the release parameters $b$ and $\tau$ as a function of the inverse porosity $\zeta = \lambda^{-1}$, which will be denominated as \textit{membrane content} hereafter\footnote{It should be possible to define the membrane content in various ways. For instance, one could define it as the logarithm  $\zeta = \ln \lambda^{-1}$, and even to normalize it by the maximum value. We decide to keep it as simple as possible, resembling the observed scaling behavior.}${}^{,}$\footnote{The membrane fraction is defined as $\lambda_m = 1- \lambda$ and, for small values it is related to the membrane content by:  $$\zeta = 1/\lambda = 1/(1-\lambda_m) \approx 1+\lambda_m.$$}. A detailed investigation on 3D systems without membrane ({\it i.e.}, $\zeta=1$) was done by adjusting the Weibull equation to release data coming from MC simulation and from the analytical solution of a continuous system with similar boundary conditions and will be discussed in the next subsection.

Before proceeding it should be important to note that the standard deviation $\sigma(t)$ is not commonly used or discussed in computer simulations of lattice models for drug release\cite{ gomesfilho2016, kosmidis2003a, kosmidis2007}. In this work $\sigma(t)$ was used as a data fluctuation resembling the deviations observed in real systems, resulting in an improved nonlinear fitting: in the case of a 3D capsule without membrane ($\zeta=1$) it was found that, although it does not introduce any qualitative change on the parameters $b$ and $\tau$ its use allowed an improvement on the fit to MC release data, as measured by the variance
of  residuals~\footnote{The goodness of the fit can be measured by a  value of variance of residuals $\theta = \chi^2/\nu$, where $\chi$ are the residuals~\cite{gnuplot} and $\nu$ is the difference between number of points and the number of parameters. In the case of a 3D release capsule of size $L=40$ without membrane we observed that $\theta$ decreased by three orders of magnitude when fitting the data with $\sigma(t)$, with values  $88.600$ with $\sigma$ and $\theta \approx 121 \times 10^3$ without it.}.

\subsection{Capsules without membrane\label{sub:nullmembrane}}

In the case of 3D devices without membrane ($\lambda=1$) it is possible to compare numerical simulations with the analytical solution to the diffusion equation, presented on~\ref{appendixA}. On Fig.~\ref{fig:theory-fit} release data from Eq.~(\ref{eq:solution}) and results from simulations of devices with various sizes are shown in parallel, with the Weibull function adjusted to data from the analytical/continuous model. Variables are normalized and drug release profiles are shown in (a) regular and (b) triple log~\footnote{In the current context, a triple log corresponds to a double logarithm in the vertical axis and single logarithm in the horizantal axis.} perspectives. As discussed previously~\cite{gomesfilho2016}, in the regular perspective Weibull functions seems to accurately fit release data in the whole time interval. Nevertheless, such analysis can be misleading and a proper way to visually assess the quality of the fitting procedure is to plot the data in a triple-log graph where the Weibull function draws a straight line with slope $b$. This point has as discussed by Langenbucher~\cite{langenbucher1972} and others~\cite{ignacio2017}. As depicted in Fig.~\ref{fig:theory-fit}~(b), the release data slight deviates from the straight line on initial and final release times, indicating that both regimes are not described with precision by this empirical function. Ignacio and Slater noted that the Weibull function presents an average compromise between the two limits and proposed a new, physically compatible, semi-empirical function that would be suited for adjusting those limiting conditions~\cite{ignacio2017}. In Ref.~\cite{christidi2016} Christidi and Kalosakas have shown that simulation results deviates from Weibull mostly in the final release, where the device loses its pharmaceutical usage, thus justifying its usage on the context of lattice models for drug release. A more detailed analysis on the data from Fig.~\ref{fig:theory-fit} allows us to infer that, in our model, MC data only starts to significantly deviate from the adjusted Weibull function after $95\%$ of the capsule drug content has being released, as observed before for another lattice model~\cite{christidi2016}.

\subsection{Phenomenological relation between $\tau$ and the membrane content.}

\begin{figure}[t]
    \begin{center}
        \includegraphics[scale=1.05]{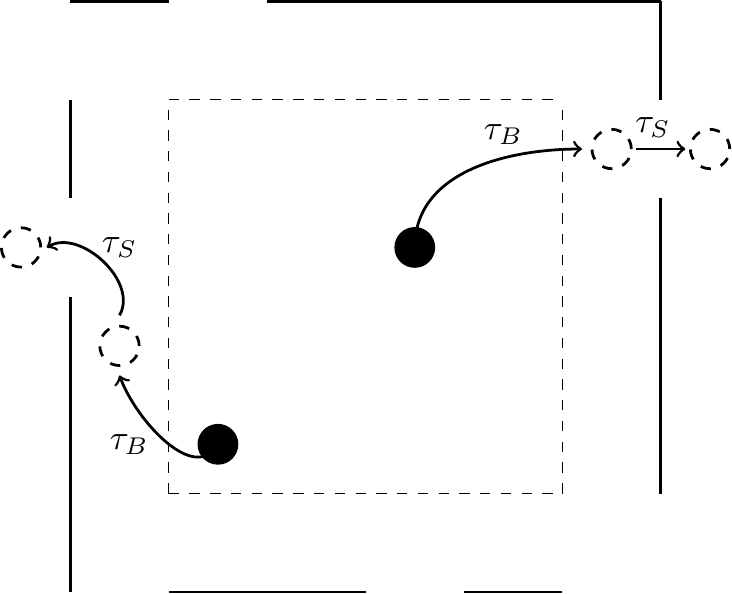}
        \includegraphics[scale=.68]{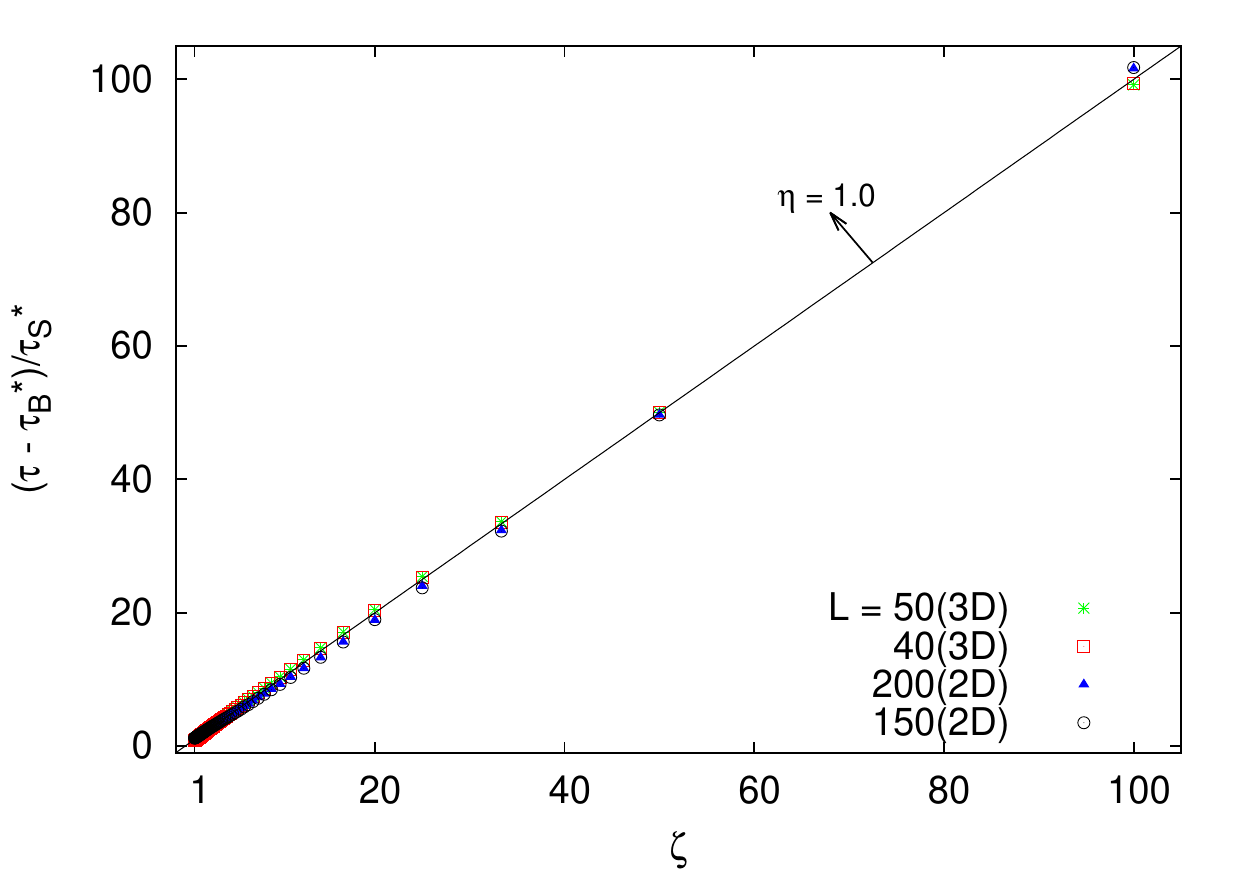} 	
        \caption{\label{fig:tau}(top) Phenomenological approach used to obtain the dependence of the average residence time inside the capsule as a function of the porosity, Eq.~(\ref{eq:tau_zeta}). Bulk drug molecules spend a time $\tau_B$ to leave the bulk, reaching the inner surface near to the membrane, and a time $\tau_S$ to be released from the surface. (bottom) The normalized characteristic time, $(\tau - \tau^*_B)/\tau^*_S$, as a function of membrane content, $\zeta=\lambda^{-1}$ (see text), for different capsule sizes in 2D and 3D lattice models.}        
    \end{center}
\end{figure}

%Table with adjusted parameters
\begin{table*}[h] 
\centering
  \begin{tabular}{c c c c c c c}
  \hline
  \hline
Model       &   L    & $\tau_B^{*}$    &  $\tau_S^{*}$ &  $\tau_B^{**}$&  $\tau_S^{**}$ & $\eta$      \\
  \hline
    $2D$    & $150$  &  $2195\pm17$ & $394\pm1$   & $2383\pm5$    & $309\pm1$    & $1.055(1)$ \\ 
	        & $200$  &  $4087\pm20$ & $533\pm2$   & $4315\pm9$    & $430\pm3$    & $1.049(1)$ \\ 
  \hline
   $3D$	    & $40$   & $134\pm1$   &  $55.3\pm0.1$   & $120.6\pm0.7$ & $61.5\pm0.3$ & $0.975(1)$ \\ 
	        & $50$   & $215\pm2$   &  $69.3\pm0.1$   & $194.6\pm0.9$ & $79.2\pm0.4$ & $0.969(1)$ \\  
   \hline
  \hline
 \end{tabular}
  \caption{ Parameters relating the characteristic Weibull time $\tau$ and membrane content, from expressions (\ref{eq:tau_zeta}) and (\ref{eq:tau_zeta_eta}), fitted to   drug release data from model devices in two and three dimensions.}
  \label{tab:coefficients}
\end{table*}

Now we will derive a phenomenological relation between a typical release and the porosity $\lambda$ (or membrane content, $\zeta = \lambda^{-1}$), which is illustrated on Fig.~\ref{fig:tau} (top). For a large capsule only a small fraction of the particles within it are located near the membrane and the average release time can be approximately described by the average release time of a typical bulk particle. Without considering reflection on the membrane any bulk particle will spend an average time $\tau_{B}$ before reaching the interior surface of the capsule, and an additional average time $\tau_{S}$ to be released from the neighbourhood of the membrane to the exterior medium. Since bulk sites are physically separated from the membrane pores by interior surface sites it should be reasonable to think that $\tau_{B}$ is dependent on the capsule size but not on the number membrane pores. On the other hand, the average time that a given particle spends on a surface site near the membrane must be a function of the number of pores. It is also expected a divergence on the residence time $\tau_{S}$ in the case of a capsule completely covered by a membrane, $\lim_{\lambda \rightarrow 0} \tau_{S} \rightarrow \infty$, and its value should decrease continuously by increasing the number of pore sites, tending to a minimum value in the case of a capsule without membrane  ($\lambda=1$). Thus, it should be interesting to consider the latter function in terms of membrane content, $\zeta = \lambda^{-1}$, and expand it as:
\begin{equation}
    \tau_{S}(\zeta) = \tau_{S}(0) + \tau'_S(0) \zeta + \mathcal O (\zeta^2).
\end{equation}
Using this expression on the Weibull parameter $\tau$ one obtains:
\begin{equation}
    \tau \approx \tau^*_B + \tau^*_S(0)\zeta, \label{eq:tau_zeta}
\end{equation}
where $\tau^*_B = \tau_B + \tau_{S}(0)$ and $\tau^*_S = \tau'_S(0)$.

Despite the simplicity of the arguments leading to expression~(\ref{eq:tau_zeta}), it reasonably reproduces the behavior of $\tau(\zeta)$ for 2D and 3D capsules under all porosity regimes, with $\tau^*_B$ and $\tau^*_S$ as parameters adjusted from simulations, as shown in Fig.~\ref{fig:tau} (bottom). The use of a second order approximation in~(\ref{eq:tau_zeta}) resulted in corrections three orders of magnitude smaller than those of the first order contribution, indicating that eq.~(\ref{eq:tau_zeta}) is a reasonable approximation to $\tau(\zeta)$. Since the curves in this figure were almost linear, we also considered the surface residence time scaling as a power law of the membrane content, as
\begin{equation}
    \tau \approx \tau^{**}_B + \tau^{**}_S(0)\zeta^\eta, \label{eq:tau_zeta_eta}
\end{equation}
where $\eta$ is a real coefficient. Table~\ref{tab:coefficients} lists the values of parameters appearing on Eqs.~(\ref{eq:tau_zeta_eta}) and~(\ref{eq:tau_zeta}) as well as those used to investigate the behavior of $b$ (discussed in the following subsection). It is interesting to stress that the values of $\eta$, which should be used as a scaling coefficient describing a scaling behavior connecting $\tau$ to $\zeta$, eq.~(\ref{eq:tau_zeta_eta}), is close to identity, providing another validation for the phenomenological derivation leading to eq.~(\ref{eq:tau_zeta}).

\subsection{Scaling behavior of $b$ with membrane content}

\begin{figure*}
  \begin{center}
    \includegraphics[scale=.68]{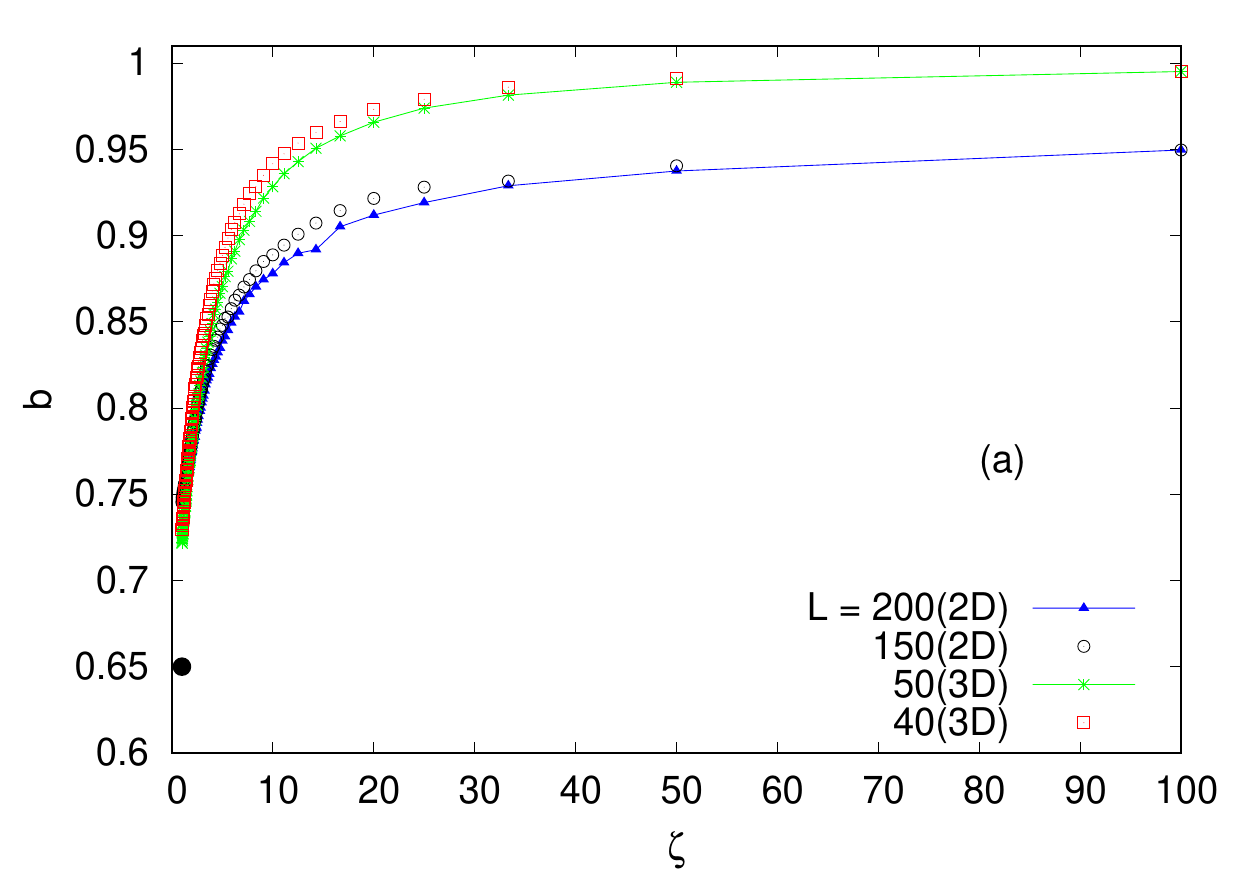}
    \includegraphics[scale=.68]{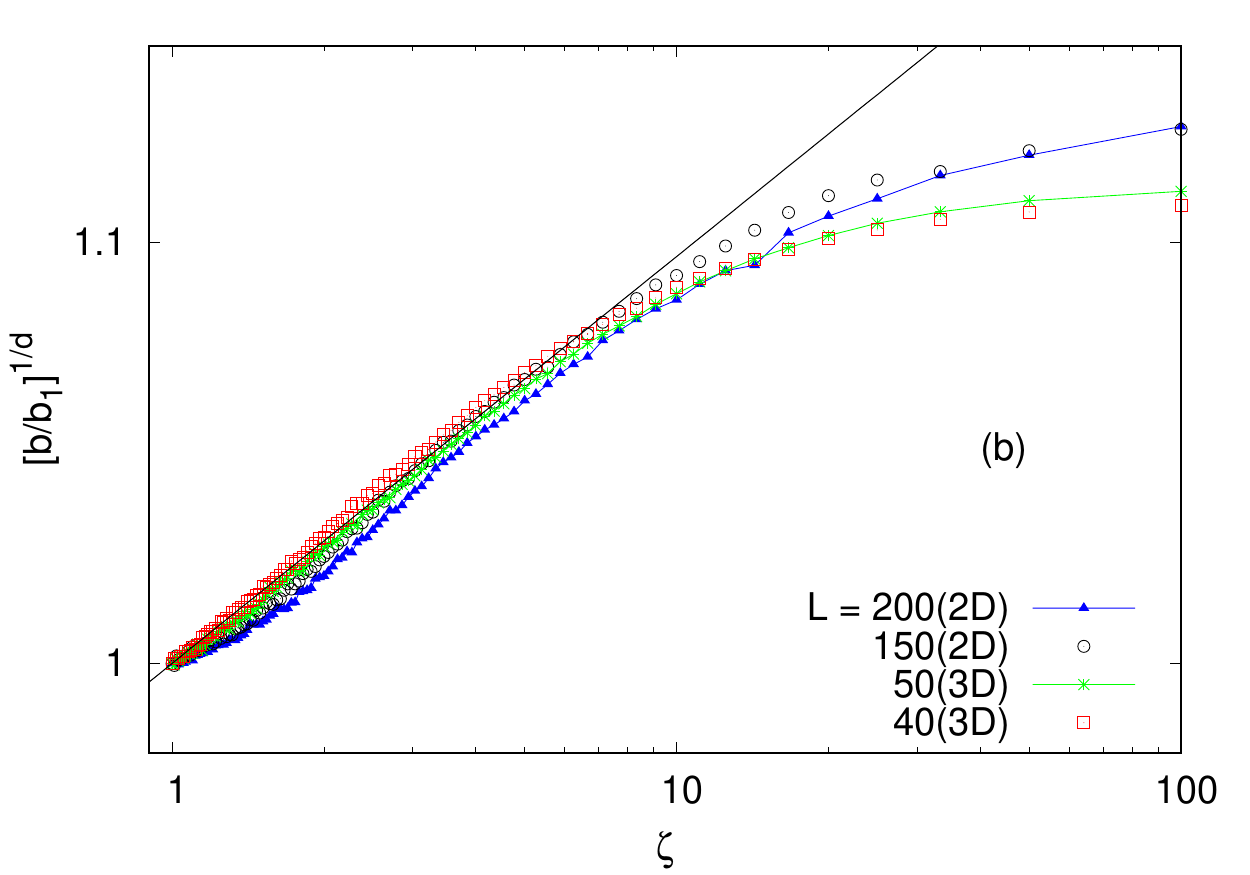} 	
    \caption{\label{fig:b}(a) Semi-empirical release parameter $b$ against the membrane content. The point (\textbullet) denotes the value of $b' \approx 0.65$, obtained from the solution to the diffusion of a device without membrane. (b) Log-log plot of the release parameter $b$ normalized, $[b/b_1]^{1/d}$,  as a function of  $\zeta$, where $d$ is the capsule dimensionality.}
  \end{center}
\end{figure*}

In Figure~\ref{fig:b} (a) we show the  semi-empirical Weibull  parameter $b$ as a function of the membrane content $\zeta$ for 2D (3D) capsules with sizes $L = 150$ and $200$  ($L=40$ and $50$).  A single filled point (\textbullet) indicates the value $b \approx 0.65$ obtained from the solution of the diffusion equation without membrane, whose diffusion profile is shown in Fig.~\ref{fig:theory-fit}. Note that the same trend is observed in 2D and 3D:  as a blocking membrane starts to cover the capsule $b$ increases as a power law from $b (\zeta = 1) \approx 0.75$  up to a point where 90\% capsule is covered with a membrane (at about $\lambda^{-1}=10$) and,  after this region, $b$ values start to increase in a less pronounced way. In two dimensions, increments in $b$ with membrane content are slower but it still converges to an uncorrelated  exponential decay, with $b \approx 1$, as found in our previous work (see Ref.~\cite{gomesfilho2016}). The same tendency is observed regardless the sizes or dimensions. 
The behavior of the Weibull parameter $b$ for values of membrane content $\zeta$ corresponding from no membrane ($\zeta=1$) to about $90\%$ of membrane coverage can be adjusted with the power law expression:
\begin{equation}
  b = b_1\zeta^{\mu},\label{eq:b-scaling}
\end{equation}
where $b_1 \equiv b(\zeta = 1)$ is the size dependent contribution to $b$, and $\mu$ is a scaling factor. In principle, the empirical coefficient $b$ does depend on size, porosity (or membrane content) and system dimension. Since the contributions coming from size and porosity were already taken account on expression~(\ref{eq:b-scaling}) through $b_1$ and $\zeta$, one could expect the scaling coefficient $\mu$ to depend only on the system dimension, $d$. This hypothesis is supported by Fig.~\ref{fig:b} (b), where it is possible to infer that $\mu$ can be adjusted with a linear relation of the type
\begin{equation}
  \mu \approx a \times d
\end{equation}
for capsules with less than $90\%$ of membrane content, with $a \approx 1/25$. Note that Fig.~\ref{fig:b} (b) confirms that a linear relation for $\ln \zeta \approx \ln (b/b_1)^{1/d}$ is approximately valid in the interval $1 \le \zeta < 10$. 

%The fact that this relation is valid up to $90\%$ of membrane content is interesting, from the experimental point of view, since it should be harder to control the device porosity using for instance, deposition methods, which are intrinsically randomized processes, for low porosity values~\ref{bla-experimental}.

\section{\label{sec:con} Conclusions}

A simple lattice, statistical mechanical model based on the lattice gas, has been proposed to investigate the dependence of the drug release profile and porosity in 2D and 3D models. The number of drug particles inside the device was adjusted to the semi-empirical Weibull function, and a detailed investigation on the dependence of the semi-empirical parameters $b$ and $\tau$ with porosity (or membrane content) was performed, based on Monte Carlo simulations. While a simple phenomenological argument allowed us to write a linear relation between the characteristic time $\tau$ and the membrane content, here defined as $\zeta = 1/\lambda$, numerical results allowed us to write $b$ in terms of a power law with a scaling exponent proportional to the system dimension. Both results were validated using well converged Monte Carlo simulations, where both the number of particles and its  standard deviation across multiple simulations were collected, with the latter being used as a fluctuation whose usage significantly improved the fitting to numerical data.

It should be interesting to note that our results corroborates the classification of $b$ values proposed by Papadopoulou and {\it et~al.}~\cite{papadopoulou2006}, which estimated that for $b < 0.75$ the release mechanism is mainly dominated by Fickian diffusion, while diffusion is combined with another mechanism for values in the range $0.75 < b <  1.0$~\cite{papadopoulou2006}. 
This is in accordance with our results from Fig.~\ref{fig:b} (a), since for membrane content values corresponding to few membrane sites, \textit{i.e.},  $\zeta \rightarrow 1$, we found $b$ in the range $0.65-0.75$, indicating that random displacements, or Fickian diffusion, is the principal mechanism for drug particles to release the capsule. On the other hand, when the number of membrane sites are increasing, and thus the escape probability for particles inside the device are decreasing (high membrane content values in the $x-$axis of Fig.~\ref{fig:b} (a)), $b$ values are increasing in the range between $0.75$ and $1.00$. Again, this is in agreement with the Papadopoulou~\textit{et~al.} criteria~\cite{papadopoulou2006} since, in this regime, the heterogeneous membrane becomes a major factor on limiting diffusion, thus blocking drug particles from being released to the outside environment. Concomitant to the increase on $b$, observed with the increase on membrane content.

We expect that our results can be useful for estimating release patterns in real drug carriers, such as porous silicon nanoparticles, were porosity can be controlled by fabrication and/or surface functionalization. It would be possible, for instance, to use Eqs.(\ref{eq:tau_zeta}) and (\ref{eq:b-scaling}) to extrapolate the release profile of a small number of samples in order to obtain the desired release profile, on monolitic devices covered with porous membranes, thus reducing the number intermediate sample devices that should be prepared.

\section{Acknowledgments}

MSGF thanks Eugene Terentjev for useful discussions. This work has been supported by CNPq, FAPDF and CAPES.

\appendix
\section{\label{appendixA} Drug release from the solution of the diffusion equation solution for 3D capsules without membrane}

We use the diffusion equation to investigate a three-dimensional cubic capsule of size $L$, with constant initial drug load in analogy to our 3D lattice model without membrane. By assuming that the  diffusion coefficient $D_0$ is constant inside the capsule the Fick's  second law may be written as~\cite{crank, siepmann2012}:
\begin{equation}
  \frac{\partial C}{\partial t} = D_0  \nabla^2 C,
  \label{eq:eq_governo}
\end{equation}
where  $C \equiv C(\mathbf{x},t)$  is the drug concentration as a function of  the position  $\mathbf{x}$ and the time  $t$. The above equation is subject to the initial condition and the boundary condition:
\begin{eqnarray}
 C(\mathbf{x}, 0) & = & C_{0},  \label{eq:ci}  \\
 C(0, t) & = &  C(L, t) = 0,    \label{eq:cc}
\end{eqnarray}
where  $C_0 = N_0/L^3$ is the initial drug concentration and $N_0$ is the initial number of particles into the device.

The equations~(\ref{eq:eq_governo}),~(\ref{eq:ci}) and (\ref{eq:cc}), can be solved by standard variable separations and power series techniques. 
By introducing reduced units:
\begin{eqnarray}
   \left\{ \begin{array}{c}
       x^* = x/L, \hspace*{0.5cm} y^* = y/L, \hspace*{0.5cm} z^* = z/L, \vspace*{0.2cm} \\
       t^* = D_0t/L^2, \vspace*{0.2cm} \\ . 
       
      C^{*} ( x^{*}, y^{*}, z^{*}, t^{*} ) = \dfrac{C (x,y,z, t)}{C^3_0},
    \end{array} \right.
    \label{eq:reduced}
\end{eqnarray}
the solution becomes:
\begin{equation}
    C^{*}( x^{*}, y^{*}, z^{*}, t^{*}) =  \frac{64}{\pi^3} \sum_{i,j,k=0}^{\infty}  \frac{C_{ijk} \exp \{-\pi^2 t^* A_{ijk}  \}}{ B_{ijk}}  \label{eq:concentration}     
\end{equation}
where $A_{ijk} = [(2i+1)^2 + (2j+1)^2 + (2k+1)^2]$, $B_{ijk} = (2i+1)(2j+1)(2k+1)$ and $C_{ijk} = \sin[(2i+1)\pi x^*]  \sin[(2j+1)\pi y^*] \sin[(2k+1)\pi z^*]$. 
The number of particles inside the capsule can be calculated as:
\begin{equation}
    \begin{split}
      N^*(t^*) &=  \int_{0}^{1} dx^* \int_{0}^{1} dy^* \int_{0}^{1} dz^* C^{*} ( x^{*}, y^{*}, z^{*}, t^{*} )\\
	           &=  \frac{512}{\pi^6}  \sum_{i,j,k=0}^{\infty}   \frac{\exp \{-\pi^2 t^* A_{ijk}  \}}{ B_{ijk}^2}. \label{eq:solution}
    \end{split}
\end{equation}
 For long time periods,  $t^* \gg 3\pi^{2}$, the release mechanism is described by a single exponential decay:  
\begin{equation}
    N^*(t^*) \approx  \frac{512}{\pi^6} \exp(-3\pi^2t^*).
    \label{eq:longtimeD}
\end{equation}

The average of the residence time in the capsule can be obtained from~(\ref{eq:solution}) as
\begin{equation}
 \overline{t^*} = \frac{512}{\pi^8}  \sum_{i=0}^{\infty}\sum_{j=0}^{\infty}\sum_{k=0}^{\infty} \frac{1}{A_{i,j,k} B_{i,j,k}^2},\label{eq:time-residence}
\end{equation}
The quantity $\overline{t^*}  \approx 0.0185$ can considered as a first approximation to the characteristic time $\tau$ on the Weibull function.

\bibliography{references}
  
\end{document}